\documentstyle[aps]{revtex}
\begin{document}
% REVTEX dialect  Version number LQ5040
\draft
\title{ Microscopic mass formulae}
\author{J. Duflo$^a$ and A.P. Zuker$^b$}
\address{(a) Centre de Spectrom\'etrie Nucl\'eaire et de Spectrom\'etrie
de Masse (IN2P3-CNRS) 91405 Orsay Campus, France}
\address{(b) Physique Th\'eorique, CRN, IN2P3-CNRS/Universit\'e Louis
Pasteur BP 20, F-67037 Strasbourg Cedex, France}
%date{\today}
\maketitle
\begin{abstract}
By assuming the existence of a pseudopotential smooth enough to do
Hartree-Fock variations and good enough to describe nuclear structure,
we construct mass formulae that rely on general scaling arguments and on
a schematic reading of shell model calculations. Fits to 1751 known
binding energies for N,Z$\geq 8$ lead to RMS errors of 614 keV with 14
parameters and 388 keV with 28 parameters. The latter is easily reduced
to a 20 parameter form at 423 keV.
\end{abstract}
Calculations of nuclear masses reflect the traditional cleavage between
local methods of shell model origin and global ones, in which
semiclassical arguments play an important role. Out of the ten
contributions to the 1986-1987 mass predictions \cite{Haustein}, two are
simply numerical, three follow Garvey-Kelson lines, and five incorporate
a Bethe-Weizs\"acker liquid drop (LD) formula as basic ingredient. The
earlier work of Liran and Zeldes \cite{LZ} is a shell model approach and
the recent Thomas-Fermi (ETFSI) calculations of the Montr\'eal-Brussels
group \cite{Pears} are definitely global but belong to a special
category in that the number of parameters is truly small (9) and the LD
form comes as output.

At present, only the local formulae of Masson and J\"{a}necke
 \cite{Haustein} can go below RMS errors of 400 keV but they need
hundreds of parameters and cannot reach safely the drip-lines as
demanded by calculations of r-processes nucleosynthesis. The droplet
(FRDM) \cite{Haustein,MolH},and ETFSI \cite{Pears} mass tables are
designed
to extrapolate efficiently but they differ in the way they do it, and
with RMS errors of some 700 keV, are not sufficiently precise for the
more terrestrial uses.

To obtain some improvements we propose to abandon the local-global
dichotomy and go back to fundamentals. The only assumption we
shall make is that there exits a nucleon-nucleon ($NN$) pseudopotential
smooth enough to do Hartree-Fock (HF) calculations and good enough
to explain nuclear structure. The pseudopotential cannot be identical
to - although we expect it to resemble - the bare $NN$ interaction.
A formal consequence of the assumption is that we can separate the
Hamiltonian  as ${\cal H}={\cal H}_m+{\cal H}_M $ . The monopole
contribution
 ${\cal H}_m={\cal H}_m^d+{\cal H}_m' $ contains a diagonal part
\begin{equation}
  {\cal H}_m^d=\sum_{k,l} a_{kl} m_{k}(m_{l}-{\delta}_{kl})
+b_{kl}(t_{k}.t_{l}-{3m_{k}\over4}{\delta}_{kl})
\end{equation}
- written in terms of number($m$) and isospin
operators($t$) for the orbits $k,l$- whose expectation value for any
state is the average energy of the configuration to which it belongs.
(A configuration is a set of states with fixed $m$ and $t$ for each
orbit). The non diagonal term ${\cal H}_{m}'$ is such that ${\cal H}_m$
is invariant under unitary transformations. Therefore HF variation of
${\cal H}$ amounts to HF variation of
${\cal H}_m$ for closed shell and single particle (or hole) states
built on them, since for this set (which we call $cs\pm 1$) each
configuration contains a single member.

The multipole Hamiltonian ${\cal H}_M$ contains all other terms
(pairing,quadrupole, etc). It is therefore responsible for
correlations, while ${\cal H}_m$ is in charge of saturation properties.

The ${\cal H}_{m}+{\cal H}_M$ separation is a rigorous result
whose proof is sketched in \cite{ZU} and given in general in
\cite{DufourZ}.
It was used in \cite{ACZ} to demonstrate that once ${\cal H}_m$
is treated phenomenologically, a parameter-free ${\cal H}_M$ derived from
realistic interactions is sufficient to provide high quality shell model
spectroscopy in all regions where exact diagonalizations are feasible
(RMS errors below 300 keV in the $p$ and $sd$ shells, below 200 keV in
the N = 50,82 isotones and Z = 28,50 isotopes).

In shell model calculations,the full $\cal H$ is not used directly but
replaced by a renormalized version $ H=H_{m}+H_M $ adapted to finite
spaces in regions bounded by magic closures. Then $H_M$ is taken to
reproduce exactly the corresponding $cs\pm 1$ states.

Although our ultimate aim would be to discover ${\cal H}_m$, in this
paper we propose a simpler, and probably necessary first step which is
the determination of the gross features of $H_m$ by including it as
basic ingredient in a mass formula.
Let us assume then that we have some $H=H_{m}+H_M$, ready for shell
model calculations :
\begin{equation}
 H = \sum A_{kl}m_{k}m_{l} + B_{kl} t_{k}.t_{l} + H_M
\end{equation}
which is taken to be a strictly two body force by including in it the
kinetic energy after elimination of the center of mass contribution.
However, for simplicity we have omitted in eq.(2) the counter terms
in ${\delta}_{kl}$ in eq.(1) (we have checked that their effect is
negligible in the fits).

To obtain some clues about $H_m$ we reduce it to separable form  by
diagonalising the $A_{kl}$ and $B_{kl}$ matrices.

For $A_{kl}$ we have
\begin{equation}
\sum_{k,l}A_{kl} m_{k}m_{l} = \sum_{\mu} e_{\mu}({\sum}_{k}m_{k}
f_{k_\mu})^2
\end{equation}
and borrow from \cite{DufourZ} the result that a realistic force
produces
a strongly dominant $e_0$, whose eigenvector $M$ has amplitudes
\begin{equation}
   f_{k_0} =\big[(p+1)(p+2)\big]^{-1/2} ={D_p}^{-1/2}
\end{equation}
where $k$ belongs to the $p$-th harmonic
oscillator shell of degeneracy $D_p$ .

Let us study this term.
Setting $m_{p}=n_{p}+z_{p} $, where $n$, $z$ are number operators for
neutrons($\nu$) and protons ($\pi$), filling shells $p_{\nu}$ and
$p_{\pi}$ up to some $p_{f\nu}$ and $p_{f\pi}$ Fermi level, we find :
\begin{equation}
<M>=\sum_{p}{m_{p}\over\sqrt{D_p}}\cong{1\over2}[(3N)^{2/3}+(3Z)^{2/3}]
\end{equation}
where we have approximated $\sqrt{(p+1)(p+2)}\approx p+3/2 $, and used
 $N = \sum n_{p_\nu} = \sum (p_{\nu}+1)(p_{\nu}+2)\cong {1\over3}
(p_{f_{\nu}}+2)^3$ and $Z= {1\over3}(p_{f_\pi}+2)^3$.

The eigenvalue $e_0$ must behave as a typical two body matrix element,
which for a realistic force goes as
\begin{equation}
 V(\omega)_{klmn}\cong {\omega\over{\omega}_0}
V({\omega}_0)_{klmn}+O({\omega}^{2})
\end{equation}
a result from ref.\onlinecite{ACZ},but adding an $O({\omega}^{2})$
correction
warranted for large oscillator constant $\omega$. Then we know that
\begin{equation}
 \hbar\omega={34.6A^{1/3}\over <r^{2}>}\cong 40A^{-1/3}+O(A^{-2/3})
\end{equation}
where the leading term is the classical result \cite{BM} obtained with a
standard mean  square radius $<r^{2}>=0.86A^{2/3}$. The $O(A^{-2/3})$
correction comes because the light nuclei are larger than the standard
estimate. By combining eqs.(6) and (7) we obtain the scaling law
$ \Gamma(A)=\Gamma A^{-1/3}+\gamma A^{-2/3} $, or its more flexible
generalization
\begin{equation}
   \Gamma(A)=(\Gamma/R)(1-\rho(\Gamma)/R))\quad,\quad R=R_c^2/A^{1/3}
\end{equation}
that we shall use for all amplitudes $\Gamma(A)$ affecting operators
${\hat\Gamma}$. The form of $R$ is left free to allow for the better
estimate:$\quad<r^2>=0.90 R_c^2,\quad$with $R_c=A^{1/3}(1-\xi\,
(2T/A)^2)^{1/3},\quad \xi=0.42$ \cite{DU}.

For the leading monopole term we have from eqs.(5) and (8)
\[{\hat\Gamma}\Gamma(A)=(\sum m_p/\sqrt{D_p})^2 e_0=O(A)+O(A^{2/3}).\]

This remarkable object goes asymptotically as volume plus surface LD
terms and at the same time produces strong magicity at the harmonic
oscillator (HO) closures, as can be checked by plotting $\hat\Gamma$
(or more precisely its expectation value ). To obtain the usually
observed extruder-intruder (EI) closures we have to add spin-orbit (SO)
effects and we rely on the diagonal construction (3) to propose a term
orthogonal to $M$ of the form $ S=\sum S_p{\cal N}_p^{-1} $, where
${\cal N}_p$ is a normalization to be determined and
\begin{equation}
S_p=pm_{jp}-2m_{rp}=(\widetilde{l.s})_p.
\end{equation}
Here (refer to Fig.1) $jp$ is the largest orbit in the $p$-th shell
and $rp$ regroups all the others. $(\widetilde{l.s})_p$ is the operator
 that produces the same splittings as $(l.s)_p$ and then collapses the
$r$-orbits to their centroid value. The rationale for considering only
two types of orbits is clear from Fig.1 : we want to give top priority to
shell formation (i.e. the $cs$ part of the $cs +1$ set ).The combinations
of $m_k$ operators other than $m_j$ and $m_r$ will contribute to subshell
 effects that we incorporate in $H_M$ and treat later.

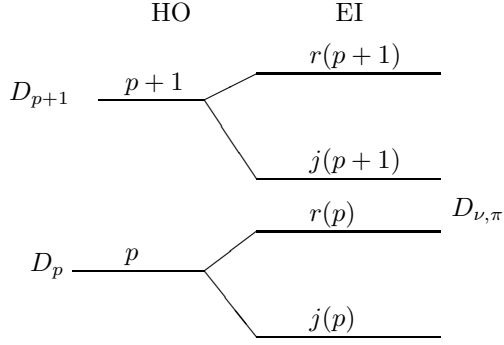
\begin{figure}
\setlength\unitlength{0.7mm}
\begin{picture}(90,90)
\protect
\put(55,90){HO}
\protect
\put(90,90){EI}
\protect
\put(28,75){$D_{p+1}$}
\protect
\put(50,77){$p+1$}
\protect
\put(85,82){$r(p+1)$}
\protect
\put(45,75){\line(1,0){20}}
\protect
\put(65,75){\line(2,1){10}}
\protect
\put(65,75){\line(2,-3){10}}
\protect
\put(75,80){\line(1,0){35}}
\protect
\put(85,62){$j(p+1)$}
\protect
\put(75,60){\line(1,0){35}}
\protect
\put(112,53){$D_{\nu ,\pi}$}
\protect
\put(85,52){$r(p)$}
\protect
\put(32,42.5){$D_p$}
\protect
\put(50,44.5){$p$}
\protect
\put(40,42.5){\line(1,0){25}}
\protect
\put(65,42.5){\line(4,3){10}}
\protect
\put(65,42.5){\line(4,-5){10}}
\protect
\put(75,50){\line(1,0){35}}
\protect
\put(85,32){$j(p)$}
\protect
\put(75,30){\line(1,0){35}}
\protect
\put(0,20){$D_{jp}=2(p+1)\quad \quad D_{rp}=p(p+1)$}
\put(0,10){$ D_{\nu,\pi}=D_p+2 =D_v
\quad\quad n_{\nu,\pi}=n_v=n_{j(p+1)}+n_{rp}$}
\protect
\caption{ HO and EI major shells.}
\protect
\end{picture}
\end{figure}
Since $M$ and $S$ are symmetric combinations of (properly scaled !) $m_p$
and $S_p$ operators, the only other contributions we can include
consistently must be symmetric in $p$, i.e., sums of $m_p^2$ and $s_p^2$,
 again properly scaled. The arguments are exactly the same for the $t$
operators.

Calling $\quad m_p=n_p+z_p,\quad t_p=\vert n_p-z_p\vert$,
$\quad S_p=p(n_{jp}+z_{jp}-2(n_{rp}+z_{rp})$ , $\quad$ and$\quad$
$St_p=p\vert n_{jp}-z_{jp}\vert -2\vert n_{rp}-z_{rp}\vert$ ,

we introduce the variables
\begin{mathletters}
\begin{equation}
  MA_p ={ m_p \over\sqrt{D_p}}\quad,\quad SA_p={S_p\over 2(p+1)}\quad,
\end{equation}
\begin{equation}
  MT_p ={ t_p \over\sqrt{D_p}}\quad,\quad ST_p={St_p\over 2(p+1)}\quad.
\end{equation}
\end{mathletters}
In general $A=N+Z$ and $T=\vert N-Z\vert/2$, but in combinations as
above,
they mean ``isoscalar'' and ``isovector'' respectively, while $M$
stands for ``master'' and $S$ for ``spin-orbit''. The first part of Table
1 gives the 12 possible symmetric quadratics obtained with operators (10)
. $F$ here stands for ``full'' and $P$ for ``partial''. All these
operators will be affected by coefficients that scale as $\Gamma (A)$ in
eq.(8). The $FCT$ term is chosen to go as an ordinary $\zeta\, l.s$ term
with
$\zeta = O(A^{-2/3})$ \cite{BM}. The other scalings then follow by
symmetry.
It could be argued that the $S$ operators in eq.(10) should carry an
extra $O(p^{-1/2})$ factor. This uncertainty is of little consequence.
On the contrary the $D_p^\alpha$ factor is important and we shall let
the fits decide in favour of $\alpha =1/2$.
\begin{minipage}[t]{18.0cm}
\begin{table}
\caption{The operators $\hat\Gamma$ ({\it called $\Gamma$ here})
in $H_m$, $H_s$ and $H_d$ }
\begin{tabular}{lll}
\tableline
 $H_m\quad (T=\vert N-Z\vert/2)$&$\alpha=1/2$&$R_c=[A(1-\xi (2T/A)^2)]
 ^{1/3} \quad \xi=0.42$\\
 $FMA=(\sum MA_p)^2$&$FSA=(\sum SA_p)^2$&$FCA=\sum MA_pD_p^{-1/2}\sum
 SA_pD_p^{-1/2}$\\
 $PMA=\sum (MA_p)^2D_{p}^{\alpha}$&$PSA=\sum (SA_p)^2D_{p}^{\alpha}$&
 $PCA=\sum (MA_p)(SA_p)D_p^{\alpha -1}$\\
 $FMT=(\sum MT_p)^2$&$FST=(\sum ST_p)^2$&
 $FCT=\sum MT_pD_p^{-1/2}\sum ST_pD_p^{-1/2}$\\
 $PMT=\sum (MT_p)^2D_{p}^{\alpha}$&$PST=\sum (ST_p)^2D_{p}^{\alpha}$&
 $PCT=\sum (MT_p)(ST_p)D_p^{\alpha -1}$\\
 $4T(T+1)A^{-2/3}$&$V_p=-mod(N,2)-mod(Z,2)$&
 $V_c=[-Z(Z-1)+.76[Z(Z-1)]^{2/3}]/R_c$ \\
 \------------&&\\
 $H_s \quad(\bar{n}=D_{\nu}-n,\quad \bar{z}=D_{\pi}-z)$&
 $S2=n\bar{n}D_{\nu}^{-1}+z\bar{z}D_{\pi}^{-1}$&
 $S3=n\bar{n}(n-\bar{n})D_{\nu}^{2(\beta -1)}+z\bar{z}(z-\bar{z})
 D_{\pi}^{2(\beta -1)}$\\
 $SQ+=2(n\bar{n})^2D_{\nu}^{2\beta -3}+2(z\bar{z})^2D_{\pi}^{2\beta -3}$&
$SQ-=4n\bar{n}z\bar{z}(D_{\nu}D_{\pi})^{\beta -3/2}$&$\beta=1/2$\\
\------------&&\\
 $H_d \quad (n'=n-JU , \bar{n}'=\bar{n}+JU) ,$&
 $(z'=z-JU, \bar{z}'=\bar{z}+JU)$&$JU=4$\\
 $QQ_1^0=(n'\bar{n}'D_{\nu}^{-3/2}\pm z'\bar{z}'D_{\pi}^{-3/2})^2$&
 $QQ\pm =QQ0\pm QQ1$ &$DK=16.$\\
\end{tabular}
\end{table}
\end{minipage}
In addition to these terms $H_m$, includes standard pairing ($V_p$) and
Coulomb ($V_c$) contributions as well as a $4T(T+1)$ term whose presence
is necessary.

If $H_m$ has ensured shell formation, mostly of EI type, the variables
that become important in modelling configuration mixing are $n_v$ and
$z_v$, the number of valence particles in EI spaces of degeneracy $D_\nu$
and $D_\pi$ (see Fig.1). If we assume a-priori the right closures and the
 right boundaries between spherical and deformed nuclei, we now that
extremely good fits can be obtained with monopole-like forms \cite{DU} .

Let us see how in the present formulation $H_M$ is expected to provide
naturally good boundaries, once $H_m$ has provided good closures.
The second part of Table 1 lists the four possible operators  whose
appearance is guaranteed under very general circumstances for {\it
spherical} nuclei. To a large extent they account for subshell effects.
They are discussed in detail in \cite{ZU}. The $D^\beta$ factors
reflect scaling uncertainties to be resolved by the fit.

Nilsson diagrams indicate that the onset of deformation is associated
with the interruption of normal spherical filling by the promotion of
4 neutrons and 4 protons to configurations in the next HO shell (i.e.
including the $j$ orbit as well as others) \cite{ZU}. Calling $q_i$ the
quadrupole moment of these intruders and $q$ that of the orbits left
behind, we expect a gain in energy of the form
\[(q+q_i)^2 = q^2+q_i^2 +2qq_i \] .
 In the third part of Table 1 we give the
expressions for $QQ0$ and $QQ1$ representing the
two possible $q^2$ contributions. Since $q_i$
is a constant, its effects are included in the DK term mainly meant
to correct the estimate of monopole loss coming from $H_m$. The $qq_i$
terms are conceptually important since $q_i$ provides the effective
quadrupole strength that will drive the lower orbits. However, we have
left
them out of the table because the information and the few keV they bring
do not justify bothering with 4 extra parameters. Conversely $JU$ is
introduced explicitly to stress that by varying it, $JU=4$
{\it turns out to be optimal}.
 The form of $q$ is equivalent to equidistant Nilsson
orbits and it is scaled so that deformation energies have a standard
$A^{1/3}$ behaviour.

All operators (except $V_c$) are affected by factors of type (8) with
$\xi=0.42$ in $R_c$ .
Energies (taken to be positive) are given by the expectation values
\[ E(N,Z)=<H_m>+<H_s>(1-\delta_d)+<H_d>\delta_d\]
\begin{equation}
 =max(<H_m>+<H_s> , <H_m>+<H_d>)
\end{equation}
the lowest possible orbits are filled for spherical nuclei
($\delta d=0$),
 while for deformed ones ($\delta d=1$), $JU$ particles are promoted to
orbits $j$. The calculations are conducted
\begin{minipage}[t]{18.0cm}
\begin{table}
\caption{Parameters of the $14p$, $28p$, $28p^\star$ and $20p$ fits.
($V_p$ and $V_c$ given in Table 3).  $28p^\star$ fit uses $R=A^{1/3}$.}
\begin{tabular}{ldddddddddddddd}
$\hat\Gamma$&FM+&PM+&4T(T+1)&FS+&FC+&S3&DK&QQ-&PS+&PS-&FS-&FC-&SQ-&QQ+\\
\tableline
 $\Gamma 14$&9.33&-0.602&-36.08&0.44&3.27&0.45&-10.0&6.53&&&&&&\\
 $\rho(\Gamma)14$&0.78&-&1.40&-&-&4.76&4.41&-&&&&&&\\
 $\Gamma 28$&9.51&-0.79&-36.51&5.19&-15.15&0.56&-36.1&21.6
  &-0.7&-0.1&1.0&-30.01&0.4&3.6\\
 $\rho(\Gamma)28$&0.75&-&1.40&4.24&3.71&4.73&4.75&3.54
  &5.34&4.47&3.61&3.51&4.73&-\\
 $\Gamma 28^\star$&9.55&-0.82&-31.83&5.26&-26.75&0.57&-36.9&19.0
  &-0.7&-0.1&1.05&-26.75&0.4&2.6\\
 $\rho(\Gamma)28^\star$&0.75&-&1.33&4.09&3.55&4.77&4.71&2.72
  &5.21&4.24&3.38&3.59&4.79&-\\
 $\Gamma 20$&9.67&-0.95&-34.77&5.95&-20.73&0.54&-44.2&27.5
  &-0.7&-0.16&1.07&-20.73&0.4&-\\
 $\rho(\Gamma)20$&0.825&.825&1.35&3.9&3.9&4.65&4.65&2.87
  &4.65&4.65&3.9&3.9&4.65&-\\
\end{tabular}
\end{table}
\end{minipage}
by initializing $\delta d$, fitting $E(N,Z)$ in the first
equality of eq.(11) to the 1751 mass values for $N,Z\geq 8$ in the latest
compilation \cite{AudiW}, then resetting $\delta d$ through
the second part of eq.(11) and iterating until convergence.
Table 2 contains results for 3 fits with 14, 28 and 20 parameters ($14p,
 28p, 20p$), whose RMS errors are 614, 388 and 423 KeV respectively.
 The notation
for the operators is $XX\pm=XXA \pm XXT$ and we have preset $JU=4$,
$\alpha=\beta=1/2$. The $14p$ fit is special in that {\it adding  one}
parameter gains at most 50 keV, while {\it substracting one} costs 100
keV for $\rho(DK)$, 140 keV for {\it either} $FS+$ {\it or} $FC+$ and
{\it more than} 300 keV for any other choice.

In $28p$, we have added two $QQ+$ and two $SQ-$ parameters for a gain
of 60 keV, and 10 parameters associated with spin orbit effects that
bring in 150 keV. With 28 parameters, Table 2 and 3 compare the
$\xi=0.42$ ($28p$) and $\xi=0$ ($28p^\star$) choices for $R_c$.

The $20p$ fit is a variation of $28p$ in which $QQ+$ is excluded and
five groups of operators: ($FM+, PM+$), ($4T(T+1)$), ($QQ-$), ($FS\pm ,
FC\pm $), ($PS\pm , DK, SQ-$) are constrained to have a single
$\rho(\Gamma)$ per group.
\begin{minipage}[t]{18.0cm}
\begin{table}
\caption{Asymptotic forms of the fits compared with a pure LD form
($6p$). (T2 is for 4T(T+1))}
\begin{tabular}{lddddddd}
 &$A$&$-A^{2/3}$&$-T2/A$&$T2/A^{4/3}$&$V_p$&$V_c$&RMS\\
\tableline
  $6p$&15.49&17.79&28.64&40.23&5.22&.705&2.544\\
 $14p$&14.95&12.43&26.73&36.21&5.17&.696&0.614\\
 $28p$&14.95&12.14&27.26&37.51&5.21&.699&0.388\\
 $28p^\star$&14.98&12.21&28.50&40.66&5.25&.699&0.409\\
 $20p$&14.99&12.36&27.24&37.30&5.21&.699&0.423\\
\end{tabular}
\end{table}
\end{minipage}
{ \sl Subshell effects}.The $H_s$ operators are largely devoted to
mock the energy patterns generated by subshell structure \cite{ZU}.
 The $S2$
operator is easily absorbed in $H_m$ and the heaviest task goes to $S3$.
Parametrization (8) is now a convenient tool unrelated to its original
derivation (eqs.(6) and (7)). This can be detected by an anomalously
large $\rho$ value leading to a change in sign of $S3$ at $R=\rho$,
 i.e. $A\geq 100$, the region in which $j$ orbits can start filling
before $r$ orbits are full. In general, a large $\rho(\Gamma)$ indicates
that the $\hat\Gamma$ operator is adding to its specific job some
subshell
corrections (e.g. $\rho(QQ-)$ in $28p$, for a gain of 25 keV, and
$\rho(DK)$ ($=\rho(S3)$ !) for a gain of 100 keV already mentioned).
Given
that SO is the very origin of subshell structure it comes as no surprise
that S-type operators are easily contamined by large $\rho$ ratios.

{\sl LD parameters and radii}.
Table 3 compares the parameters of $6p$ (pure LD)  with
those obtained by expanding eq.(5) and similar ones for $FM$ and $PM+$:
the combinations$\quad 1.717[(FM+)+(PM+)]\quad$ and

$4T(T+1)+(0.382+1.145\xi)[(FM+)+(PM+)]$

become the asymptotic coefficients of $A$ and $4T^2$ respectively. The
factor in $\xi$ comes from the $R$ denominators. It doubles for the
$4T^2/A^{4/3}$ term. The gain
brought about by the use of a more precise form of $R_c$ is no doubt
significant as shown by the case of $28p$.

 It should be noted  that with $\alpha=1/2$ scaling,
$PM+$ is a volume term, while for $\alpha=0$
it is pure surface and $\rho(FM+)$ becomes very small, a disturbing
result that justifies the $\alpha=1/2$ choice.

It is  remarkable, also, that the radius extracted for $V_c$ in Table 3,
$r\approx 1.235 R_c$, is very close to the fitted $r\approx 1.225 R_c$
\cite{DU}.

{\sl The $FMT-4T(T+1)$ puzzle}. The presence of the $4T(T+1)$ term is
demanded by the fits below 650 keV,
while the omission of $FMT$ is possible down to the 470 keV level. The
puzzle is that it serves no purpose to treat $FMT$ as a free parameter:
the efficient combination is $FM+ =FMA+FMT$. Renormalisation effects and
the cancellation of kinetic and potential energies are the only guesses
we can propose for the emergence of $T(T+1)$ and $FM+$ as leading
operators.

{\sl Deformed nuclei}. To avoid unphysical results we allow
$\delta d=1$ in eq.(11) only when $DK > 0$, i.e.for $A \geq 100$.
  The number of
$\delta=1$ cases is (423,367,329) for the ($14p, 28p, 20p$) fits, and the
RMS errors of (594,302,385) are smaller than those for the spherical
nuclei in each case. The experimental ground state bands of the
$\delta_d =1$  nuclei show rotational features.
\vspace{0.4 cm}

{\sl To conclude}: by replacing the global-local alternative by the
monopole-multipole separation of $\cal H$ we can find good mass formulae.
Better ones will come once we learn more about ${\cal H}_m $.

\vspace{0.4 cm}
%acknowledgments
We would like to thank G. Audi, G.E. Brown, E. Caurier,
T.von Egidy, B. Jonson, P. M\"oller, H. Niefnecker, M. Pearson, A. Poves,
P. Quentin, A. Sobicewsky, F.K. Thielemann, F. Tondeur and N. Zeldes for
useful exchanges.

\end{document}